\documentstyle[11pt]{article}
\bf
 \newcommand{\hs}[1]{\hspace*{ #1 mm}}

 \newenvironment{proof}{\par \noindent
            {\bf Proof. \hs{2}}}{\hfill$\Box$ \vspace*{3mm}}

\newcommand{\ignore}[1]{}

\newcommand{\cent}{{|}\!\!\mathrm{c}}

\begin{document}
\title{{\bf Some observations on two-way finite automata with quantum and classical states }}
\author{Daowen Qiu\\
   {\small {\it Department of Computer Science, Zhongshan University, }}\\
   {\small {\it  Guangzhou, 510275, People's Republic of China}}\\
    {\small {\it E-mail address:  issqdw@mail.sysu.edu.cn}}}
\date{  }
\maketitle
\begin{center}
\begin{minipage}{130mm}

 {\bf Abstract}
 \par
\hskip 5mm  {\it Two-way finite automata with quantum and
classical states} (2qcfa's) were introduced by Ambainis and
Watrous. Though this computing model is more restricted than the
usual {\it two-way quantum finite automata} (2qfa's) first
proposed by Kondacs and Watrous, it is still more powerful than
the classical counterpart. In this note, we focus on dealing with
the operation properties of 2qcfa's. We prove that the Boolean
operations (intersection, union, and complement) and the reversal
operation of the class of languages recognized by 2qcfa's with
error probabilities are closed; as well, we verify that the
catenation operation of such class of languages is closed under
certain restricted condition. The numbers of states of these
2qcfa's for the above operations are presented. Some examples are
included, and $\{xx^{R}|x\in \{a,b\}^{*},\#_{x}(a)=\#_{x}(b)\}$ is
shown to be recognized by 2qcfa with one-sided error probability,
where $x^{R}$ is the reversal of $x$, and $\#_{x}(a)$ denotes the
$a$'s number in string $x$.

\vskip 3mm


\end{minipage}
\end{center}

\section*{1.  Introduction}

Quantum computers---the physical devices complying with quantum
mechanics were first suggested by Feynman [15] and then formalized
further by Deutch [12]. A main goal for exploring this kind of
model of computation is to clarify whether computing models built
on quantum physics can surpass classical ones in essence.
Actually, in 1990's Shor's quantum algorithm for factoring
integers in polynomial time [30] and afterwards Grover's algorithm
of searching in database of size $n$ with only $O(\sqrt{n})$
accesses [17] have successfully shown the great power of quantum
computers. Since then great attention has been given to this
intriguing field in the academic community [19,27], in which the
study of clarifying the power of some fundamental models of
quantum computation is of interest [19, pp. 151-192].

{\it Quantum finite automata} (qfa's) can be thought of
theoretical models of quantum computers with finite memory. With
the rise of exploring quantum computers, this kind of theoretical
models was firstly studied by Moore and Crutchfield [24], Kondacs
and Watrous [23], and then Ambainis and Freilds [1], Brodsky and
Pippenger [11], and the other authors (e.g., name only a few,
[2,4,5,7,8,9,10,18,25,26,28,29], and for some details we may refer
to [19]). The study of qfa's is mainly divided into two ways: one
is {\it one-way quantum finite automata} (1qfa's) whose tape heads
move one cell only to right at each evolution, and the other {\it
two-way quantum finite automata} (2qfa's), in which the tape heads
are allowed to move towards right or left, or to be stationary.
(Notably, Amano and Iwama [3] dealt with 1.5qfa's whose tape heads
are allowed to move right or to be stationary, and showed that the
emptiness problem for this restricted model is undecidable.) In
terms of the measurement times in a computation, 1qfa's have two
types: {\it measure-once} 1qfa's (MO-1qfa's) initiated by Moore
and Crutchfield [24] and {\it measure-many} 1qfa's (MM-1qfa's)
studied firstly by Kondacs and Watrous [23].

MO-1qfa's mean that at every computation there is only a
measurement at the end of computation, whereas MM-1qfa's represent
that measurement is performed at each evolution. The class of
languages recognized by MM-1qfa's with bounded error probabilities
strictly bigger than that by MO-1qfa's, but both MO-1qfa's and
MM-1qfa's recognize proper subclass of regular languages with
bounded error probabilities [1,24,11,23,7,8]. On the other hand,
the class of languages recognized by MM-1qfa's with bounded error
probabilities is not closed under the binary Boolean operations
(intersection, union, complement) [1,4,11,8], and by contrast
MO-1qfa's satisfy the closure properties of the languages
recognized with bounded error probabilities under binary Boolean
operations [11,10].

A more powerful model of quantum computation than its classical
counterpart is 2qfa's that were first studied by Kondacs and
Watrous [23]. As is well known, classical two-way finite automata
have the same power as one-way finite automata for recognizing
languages. Freivalds [16] proved that {\it two-way probabilistic
finite automata} (2pfa's) can recognize non-regular language
$L_{eq}=\{a^{n}b^{n}|n\in {\bf N}\}$ with arbitrarily small error,
but it was verified to require exponential expected time [20]. (In
this paper, ${\bf N}$ denotes the set of natural numbers.)
Furthermore, it was demonstrated that any 2pfa's recognizing
non-regular languages with bounded error probabilities need take
exponential expected time [13,22]. In 2qfa's, a sharp contrast has
arisen, as Kondacs and Watrous [23] proved that $L_{eq}$ can be
recognized by some 2qfa's with one-sided error probability in
linear time.

Recently, Ambainis and Watrous [6] proposed a different two-way
quantum computing model---{\it two-way finite automata with
quantum and classical states} (2qcfa's). In this model, there are
both quantum states and classical states, and correspondingly two
transfer functions: one specifies unitary operator or measurement
for the evolution of quantum states and the other describes the
evolution of classical part of the machine, including the
classical internal states and the tape head. Therefore, this model
can be viewed as an intermediate version between 1qfa's and
2qfa's, and it is more restricted than ordinary 2qfa's by Kondacs
and Watrous [23]. This device may be simpler to implement than
ordinary 2qfa's, since the moves of tape heads of 2qcfa's are
classical. In spite of the existing restriction, 2qcfa's have more
power than 2pfa's. Indeed, as Ambainis and Watrous [6] pointed
out, 2qcfa's clearly can recognize all regular languages with
certainty, and particularly, they [6] proved that this model can
also recognize non-regular languages $L_{eq}=\{a^{n}b^{n}|n\geq
1\}$ and palindromes $L_{pal}=\{x\in \{a,b\}^{*}|x=x^{R}\}$, where
notably the complexity for recognizing $L_{eq}$ is polynomial time
in one-sided error. As is known, no 2pfa can recognize $L_{pal}$
with bounded error in any amount of time [14]. Therefore, this is
an interesting and more practicable model of quantum computation,
and we hope to deal with further related basic properties.

Operations of finite automata are of importance [21] and also
interest in the framework of quantum computing. Our goal in this
note is to deal with the operation properties of 2qcfa's. We
investigate some closure properties of the class of languages
recognized by 2qcfa's, and we focus on the binary Boolean
operations, reversal operation, and catenation operation.
Notwithstanding, we do not know whether or not these properties
hold for the ordinary 2qfa's without any restricted condition, and
would like to propose them as an open problem (As the author is
aware, the main problem to be overcome is how to preserve the
unitarity of the constructed 2qfa's without any restricted
condition).

The remainder of the paper is organized as follows. In Section 2
we introduce the definition of 2qcfa's and related results; as
well, in terms of the results by Ambainis and Watrous [6], we
further present some non-regular languages recognized by 2qcfa's
with one-sided error probabilities in polynomial expected time.
Section 3 is the main part and deals with operation properties of
2qcfa's, including intersection, union, complement, reversal, and
catenation operations; also, we include some examples as an
application of these results derived, and we present the numbers
of states of these 2qcfa's for the above operations. Finally, some
remarks are included in Section 4.

\section*{2. Definition of 2qcfa's and some non-regular languages related}

In this section, we recall the definition of 2qcfa's, and,
introduce the 2qcfa for accepting $L_{eq}$ with one-sided error
probability in polynomial time that was verified by Ambainis and
Watrous [6].

A 2qcfa $M$ consists of a 9-tuple
\[
M=(Q,S,\Sigma,\Theta,\delta,q_{0},s_{0},S_{acc},S_{rej})
\]
where $Q$ and $S$ are finite state sets, representing quantum
states and classical states, respectively, $\Sigma$ is a finite
alphabet of input, $q_{0}\in Q$ and $s_{0}\in S$ denote
respectively the initial quantum state and classical state,
$S_{acc},S_{rej}\subseteq S$ represent the sets of accepting and
rejecting, respectively, $\Theta$ and $\delta$ are the functions
specifying the behavior of $M$ regarding quantum portion and
classical portion of the internal states, respectively.

For describing $\Theta$ and $\delta$, we further introduce related
notions. We denote $\Gamma=\Sigma\cup \{\cent, \$\}$, where
$\cent$ and $\$$ are respectively the left end-marker and right
end-marker. $l_{2}(Q)$ represents the Hilbert space with the
corresponding base identified with set $Q$. Let ${\cal
U}(l_{2}(Q))$ and ${\cal M}(l_{2}(Q))$ denote the sets of unitary
operators and orthogonal measurements over $l_{2}(Q)$,
respectively. An orthogonal measurement over $l_{2}(Q)$ is
described by a finite set $\{P_{j}\}$ of projection operators on
$l_{2}(Q)$ such that $\sum_{j}P_{j}=I$ and
$P_{i}P_{j}=\left\{\begin{array}{ll}P_{j},&
i=j,\\
O,& i\not=j,
\end{array}
\right.$ where $I$ and $O$ are identity operator and zero operator
on $l_{2}(Q)$, respectively. If a superposition state
$|\psi\rangle$ is measured by an orthogonal measurement described
by set $\{P_{j}\}$, then
\begin{enumerate}
\item the result of the measurement is $j$ with probability
$\|P_{j}|\psi\rangle\|^{2}$ for each $j$,
\item and the
superposition of the system collapses to
$P_{j}|\psi\rangle/\|P_{j}|\psi\rangle\|$ in case $j$ is the
result of measurement.
\end{enumerate}
For example, suppose $Q=\cup_{j}Q_{j}$ and $Q_{i}\cap
Q_{j}=\emptyset$ for any $i\not= j$, then all the projectors
$P_{j}$ mapping to subspaces $span \hskip 2mm Q_{j}$ spanned by
$Q_{j}$ specify an orthogonal measurement over $l_{2}(Q)$.

$\Theta$ and $\delta$ are specified as follows. $\Theta$ is a
mapping from $S\backslash (S_{acc}\cup S_{rej})\times \Gamma$ to
${\cal U}(l_{2}(Q))\cup {\cal M}(l_{2}(Q))$, and $\delta$ is a
mapping from $S\backslash (S_{acc}\cup S_{rej})\times \Gamma$ to
$S\times \{-1,0,1\}$. To be more precise, for any pair
$(s,\sigma)\in S\backslash (S_{acc}\cup S_{rej})\times \Gamma$,
\begin{enumerate}
\item if $\Theta(s,\sigma)$ is a unitary operator $U$, then $U$
performing the current superposition of quantum states evolves
into new superposition, and $\delta(s,\sigma)=(s^{'},d)\in S\times
\{-1,0,1\}$ makes the current classical state $s$ become $s^{'}$,
together with the tape head moving in terms of $d$ (moving right
one cell if $d=1$, left if $d=-1$, and being stationary if $d=0$),
for which in case $s^{'}\in S_{acc}$, the input is accepted, and
in case $s^{'}\in Q_{rej}$, the input rejected;

\item if $\Theta(s,\sigma)$ is an orthogonal measurement, then the
current quantum state, say $|\psi\rangle$, is naturally changed to
quantum state $P_{j}|\psi\rangle/ \|P_{j}|\psi\rangle\|$ with
probability $\|P_{j}|\psi\rangle\|^{2}$ in terms of the
measurement, and in this case, $\delta(s,\sigma)$ is instead a
mapping from the set of all possible results of the measurement to
$S\times \{-1,0,1\}$. For instance, for the result $j$ of
measurement, and $\delta(s,\sigma)(j)=(s_{j},d)$, then
\begin{enumerate}
\item if $s_{j}\in S\backslash (S_{acc}\cup S_{rej})$, with
probability $\|P_{j}|\psi\rangle\|^{2}$ the updated quantum state
is $P_{j}|\psi\rangle / \|P_{j}|\psi\|$ and the classical state is
$s_{j}$ together with the tape head moving by means of $d$; \item
if $s_{j}\in S_{acc}$, with probability
$\|P_{j}|\psi\rangle\|^{2}$ the machine accepts the input and the
computation halts; \item  and similarly, if $s_{j}\in S_{rej}$,
with probability $\|P_{j}|\psi\rangle\|^{2}$ the machine rejects
the input and the computation halts.
\end{enumerate}
It is seen that if the current all possible classical states are
in $S_{acc}\cup S_{rej}$, then the computation for the current
input string ends.
\end{enumerate}

On the basis of the above definition, we can naturally define the
computing process and the probabilities of accepting and
rejecting. For any input string $x\in\Sigma^{*}$, the machine
begins with the initial quantum state $|q_{0}\rangle$ and
classical state $s_{0}$ and reads the left end-marker $\cent$.
While in terms of $\Theta(s_{0},\cent)$, the quantum state is
evolved, by means of $\delta(s_{0},\cent)$ the classical state is
changed and the tape head is moved correspondingly (in accordance
with [6], the tape head is not allowed to move left (right) when
it points at $\cent$ ($\$$)). In each evolution, the corresponding
accepting and rejecting probabilities are computed in terms of
whether the transformation function $\delta$ enters accepting or
rejecting states. The computation will end if all classical states
entered are in $S_{acc}\cup S_{rej}$. Therefore, similar to the
definition of accepting and rejecting probabilities for MM-1qfa's
and 2qfa's [23], the accepting and rejecting probabilities
$P^{(M)}_{acc}(x)$ and $P^{(M)}_{rej}(x)$ in $M$ for input $x$ are
respectively the sums of all accepting probabilities and all
rejecting probabilities before the end of the machine for
computing input $x$.

A language $L$ over alphabet $\Sigma^{*}$ is called to be
recognized by 2qcfa $M$ with bounded error probability $\epsilon$
if $\epsilon \in [0,1/2)$, and \begin{itemize} \item for any $x\in
L$, $P^{(M)}_{acc}(x)\geq 1-\epsilon$, \item for any $x\in
L^{c}=\Sigma^{*}\backslash L$, $P^{(M)}_{rej}(x)\geq 1-\epsilon$.
\end{itemize}
We say that 2qcfa $M$ recognizes language $L$ over alphabet
$\Sigma$ with one-sided error $\epsilon>0$ if $P^{(M)}_{acc}(x)=1$
for $x\in L$, and $P^{(M)}_{rej}(x)\geq 1-\epsilon$ for $x\in
L^{c}=\Sigma^{*}\backslash L$.

As were shown by Ambainis and Watrous [6], for any $\epsilon >0$,
2qcfa's can recognize palindromes $L_{pal}=\{x\in
\{a,b\}^{*}|x=x^{R},\}$ and $L_{eq}=\{a^{n}b^{n}|n\in {\bf N}\}$
with one-sided error probability $\epsilon$, where $\epsilon$ can
be arbitrarily small. Here we simply describe their computing
process for recognizing $L_{eq}$, and the details are referred to
[6]. In their machine (we denote it by $M_{eq}$), there are only
two quantum states, i.e., $Q=\{q_{0}, q_{1}\}$. For any input
string $x\in \{a,b\}^{*}$, $M_{eq}$ firstly checks whether or not
$x$ is of the form $a^{n}b^{m}$ for $n,m\geq 1$. If not, the
machines rejects it immediately; otherwise, the machine reads the
input symbols from left to right successively. After reading
symbol $a$ (or $b$), the quantum state part that is described by
Hilbert space $l_{2}(Q)$ is performed by rotating unitary
transformation $U_{\alpha}$ (or $U_{\beta}$), where
$\alpha=\sqrt{2}\pi$ (and $\beta=-\sqrt{2}\pi$) is the angle
rotated. When the tape head reads the right end-marker $\$$, the
machine performs orthogonal measurement:
\begin{itemize}
\item If $\#_{x}(a)\not=\#_{x}(b)$, where $\#_{x}(a)$ (and
$\#_{x}(b)$) represents the number of $a$ (and $b$) in string $x$,
say $n$ $a$'s and $m$ $b$'s, then there is non-zero probability
(at least $\frac{1}{2(n-m)^{2}}$) for measuring $|q_{1}\rangle$.
Therefore, the machine rejects that part of $q_{1}$, and with
$q_{0}$ its tape head is moved to the first input symbol in the
left, and then by performing random walk the tape head reaches the
right end-marker $\$$, repeating this action twice and then
flipping $k$ (related to $\epsilon$) coins. If all results are not
``heads", the machine accepts with probability
$\frac{1}{2^{k}(n+m+1)^{2}}$; otherwise, with the rest probability
the machine recurs to the beginning configuration and then
executes a round again. With at most $O((n+m)^{4})$ steps, the
rejecting probability is bigger than $1-\epsilon$.

\item If $\#_{x}(a)=\#_{x}(b)=n$, then with certainty the
machine's tape head is moved to the first input symbol in the
left, and then by performing random walk the tape head reaches the
right end-marker $\$$, repeating this action twice and then
flipping $k$ (related to $\epsilon$) coins. If all results are not
``heads", the machine accepts with probability
$\frac{1}{2^{k}(n+m+1)^{2}}$; otherwise, with the rest probability
the machine recurs to the beginning configuration and then
executes a round again. With at most $O(n^{2})$ steps, the
accepting probability is bigger than
$1-(1-\frac{1}{2^{k}(n+m+1)^{2}})^{cn^{2}}$, that is close to $1$
for appropriate constant $c$.
\end{itemize}

Basing on this 2qcfa $M_{eq}$ presented above, we may further
observe that some another non-regular languages can also be
recognized by 2qcfa's with bounded error probabilities in
polynomial time, and, we would state them in the following Remarks
to conclude this section.

{\bf Remark 1.} In terms of the 2qcfa $M_{eq}$ above by Ambainis
and Watrous [6], the language
$\{a^{n}b_{1}^{n}a^{m}b_{2}^{m}|n,m\in {\bf N}\}$ can also be
recognized by some 2qcfa denoted by $M_{eq}^{(2)}$ with one-sided
error probability in polynomial time. Indeed, let $M_{eq}^{(2)}$
firstly checks whether or not the input string, say $x$, is the
form $a^{n_{1}}b_{1}^{n_{2}}a^{m_{1}}b_{2}^{m_{2}}$. If not, then
$x$ is rejected certainly; otherwise, $M_{eq}^{(2)}$ simulates
$M_{eq}$ for deciding whether or not $a^{n_{1}}b_{1}^{n_{2}}$ is
in $L_{eq}$, by using the $a$ in the right of $b_{1}$ as the right
end-marker $\$$. If not, then $x$ is rejected; otherwise, this
machine continues to simulate $M_{eq}$ for recognizing
$a^{m_{1}}b_{2}^{m_{2}}$, in which $b_{1}$ is viewed as the left
end-marker $\cent$. If it is accepted, then $x$ is also accepted;
otherwise, $x$ is rejected.

{\bf Remark 2.} For $k\in {\bf N}$, let
$L_{eq}(k,a)=\{a^{kn}b^{n}|n\in {\bf N}\}$. Obviously,
$L_{eq}(1,a)=L_{eq}$. Then, by means of the 2qcfa $M_{eq}$,
$L_{eq}(k,a)$ can be recognized by some 2qcfa, denoted by
$M_{eq}(k,a)$, with one-sided error probability in polynomial
time. Indeed, $M_{eq}(k,a)$ is derived from $M_{eq}$ by replacing
$U_{\beta}$ with $U_{\beta_{k}}$, where $\beta_{k}=\sqrt{2}k\pi$.
Likewise, denote $L_{eq}(k,b)=\{b^{kn}a^{n}|n\in {\bf N}\}$. Then
$L_{eq}(k,b)$ can be recognized by some 2qcfa $M_{eq}(k,b)$ with
one-sided error probability in polynomial time.

{\bf Remark 3.} Let $L_{=}=\{x\in
\{a,b\}^{*}|\#_{x}(a)=\#_{x}(b)\}$, where $\#_{x}(a)$ (and
$\#_{x}(b)$) represents the number of $a$ (and $b$) in string $x$.
Then $L_{=}$ is recognized by some 2qcfa, denoted by $M_{=}$, with
one-sided error probability in polynomial time. Indeed, by
observing the words in $L_{=}$, $M_{=}$ can be directly derived
from $M_{eq}$ above by omitting the beginning process for checking
whether or not the input string is of the form $a^{n}b^{m}$.

\section*{3. Operation properties of 2qcfa's}

This section deals with operation properties of 2qcfa's, and, a
number of examples as application are incorporated. For
convenience, we use notations $2QCFA_{\epsilon}(poly-time)$ and
$2QCFA(poly-time)$ to denote the classes of all languages
recognized by 2qcfa's with given error probability $\epsilon\geq
0$ and with any error probabilities in $[0,1)$, respectively,
which run in polynomial expected time; for any language $L\in
2QCFA(poly-time)$, let $QS_{L}$ and $CS_{L}$ denote respectively
the minimum numbers of quantum states and classical states of the
2qcfa that recognizes $L$ with error probability in $[0,1)$.
Firstly, we consider intersection operation.

{\bf Theorem 1.} If $L_{1}\in 2QCFA_{\epsilon_{1}}(poly-time)$,
$L_{2}\in 2QCFA_{\epsilon_{2}}(poly-time)$, then $L_{1}\cap
L_{2}\in 2QCFA_{\epsilon}(poly-time)$ with
$\epsilon=\epsilon_{1}+\epsilon_{2}-\epsilon_{1}\epsilon_{2}$.

\begin{proof} Let $M_{1}$ and $M_{2}$ be 2qcfa's for recognizing
$L_{1}$ and $L_{2}$ with error probabilities
$\epsilon_{1},\epsilon_{2}\geq 0$, respectively. The basic idea is
as follows. Firstly let the machine $M$ constructed simulate
$M_{1}$. If $M_{1}$ rejects, then $M$ also rejects; if $M_{1}$
accepts, then $M$ continues to simulate $M_{2}$ and then $M_{2}$
decides the accepting and rejecting probabilities. This 2qcfa $M$
may be more clearly described by the following process.

{\it For input string $x$, $M_{1}$ and $M_{2}$ with initial
quantum state $|q_{1,0}\rangle$ and  $|q_{2,0}\rangle$ as well as
classical state $s_{1,0}$ and $s_{2,0}$, respectively; also, $M$
has initial quantum state $|q_{1,0}\rangle$ and classical state
$s_{1,0}$. $M$ firstly simulate $M_{1}$. If $M_{1}$ rejects, then
$M$ rejects; if $M_{1}$ accepts, then $M$ becomes quantum state
$|q_{2,0}\rangle$ and classical state $s_{2,0}$, and continues to
simulate $M_{2}$. If $M_{2}$ accepts, then also $M$ accepts;
otherwise $M$ rejects as $M_{2}$ does.}

Basing on the analysis above, we now prove this theorem more
formally. Let 2qcfa's

$M_{i}=(Q_{i},S_{i},\Sigma_{i},\Theta_{i},\delta_{i},q_{i,0},s_{i,0},S_{i,acc},S_{i,rej})$\\
for accepting $L_{i}$ with error probabilities $\epsilon_{i}\geq
0$ $(i=1,2)$, where we suppose that for $i=1,2$,
\begin{itemize}
\item $Q_{i}=\{q_{i,0},q_{i,1},\ldots,q_{i,n_{i}}\}$,

\item $S_{i}=\{s_{i,0},s_{i,1},\ldots,s_{i,m_{i}}\}$.
\end{itemize}
 We construct 2qcfa
$M=(Q,S,\Sigma,\Theta,\delta,q_{0},s_{0},S_{acc},S_{rej})$ where:
\begin{itemize}
\item $\Sigma=\Sigma_{1}\cap \Sigma_{2}$, \item $q_{0}=q_{1,0}$,
\item $s_{0}=s_{1,0}$, \item $Q=Q_{1}\cup Q_{2}$ (also, we can
equivalently use $Q=Q_{1}\oplus Q_{2}$ without essential
difference), \item $S=S_{1}\cup S_{2}\cup
\{t^{(1,j)}|j=0,1,\ldots,n_{1}\}$, \item $S_{rej}=S_{1,rej}\cup
S_{2,rej}$, \item $S_{acc}=S_{acc,2}$,
\end{itemize}
and $\Theta$ and $\delta$ are defined as follows:
\begin{enumerate}
\item For any $s\in S_{1}\backslash S_{1,acc}\cup S_{1,rej}$,
$\sigma\in \Sigma\cup\{\cent,\$\}$,
\begin{enumerate}
\item if $\Theta_{1}(s,\sigma)\in {\cal U}(l_{2}(Q_{1}))$, i.e., a
unitary operator on $l_{2}(Q_{1})$, then $\Theta(s,\sigma)$ is
unitary operator on $l_{2}(Q)$ by extending $\Theta_{1}(Q)$ in
terms of $\Theta(s,\sigma)|q_{2,j}\rangle=|q_{2,j}\rangle$ for
$0\leq j\leq n_{2}$, and $\delta(s,\sigma)=\delta_{1}(s,\sigma)$;

\item if $\Theta_{1}(s,\sigma)\in {\cal M}(l_{2}(Q_{1}))$, i.e.,
an orthogonal measurement on $l_{2}(Q_{1})$, say the measurement
is specified by the set of $\{P_{j}\}$ of projectors, where each
$P_{j}$ is a projection operator and
$\delta_{1}(s,\sigma)=(s_{j},d_{j})$, then
$\delta(s,\sigma)(j)=\delta_{1}(s,\sigma)(j)$, and
$\Theta(s,\sigma)$ is an orthogonal measurement described by the
set $\{P_{j}^{'}\}\cup\{I_{2}\}$ of projectors on
$l_{2}(Q)=l_{2}(Q_{1}\cup Q_{2})$, where $P_{j}^{'}$ are
projection operators by extending $P_{j}$ with
$P_{j}^{'}|q_{2,j}\rangle=0$ for $0\leq j\leq n_{2}$, and $I_{2}$
is projection operator mapping to $l_{1}(Q_{2})$, that is, an
identity operator on $l_{2}(Q_{2})$ and $I_{2}|q_{1,j}\rangle=0$
for $0\leq j\leq n_{1}$.
\end{enumerate}

\item For any $s\in S_{1,acc}$, $\sigma\in \Sigma\cup
\{\cent,\$\}$,
\begin{enumerate}
\item if $\sigma\not=\cent$, then $\Theta(s,\sigma)=I$, where $I$
is identity operator on $l_{2}(Q)$, and $\delta(s,\sigma)=(s,-1)$;

\item if $\sigma=\cent$, then $\Theta(s,\sigma)$ is an orthogonal
measurement described by projectors $\{|q_{1,j}\rangle\langle
q_{1,j}||q_{1,j}\in Q_{1}\}$,
$\delta(s,\sigma)(1,j)=(t^{(1,j)},0)$;
$\Theta(t^{(1,j)},\cent)$=$U(q_{1,j},q_{2,0})$,
$\delta(t^{(1,j)},\cent)=(s_{2,0},0)$, where $U(q_{1,j},q_{2,0})$
is a unitary operator on $l_{2}(Q)$ satisfying
$U|q_{1,j}\rangle=|q_{2,0}\rangle$.
\end{enumerate}
\item For any $s\in S_{2}$, $\sigma\in \Sigma\cup \{\cent,\$\}$,
\begin{enumerate}
\item if $\Theta_{2}(s,\sigma)$ is a unitary operator on
$l_{2}(Q_{2})$, then $\Theta(s,\sigma)$ is a unitary operator on
$l_{2}(Q)$ by extending $\Theta_{2}(s,\sigma)$ with
$\Theta(s,\sigma)|q_{1,j}\rangle=|q_{1,j}\rangle$ for $0\leq j\leq
n_{1}$, and $\delta(s,\sigma)=\delta_{2}(s,\sigma)$;

\item if $\Theta_{2}(s,\sigma)$ is an orthogonal measurement on
$l_{2}(Q_{2})$ described by projection operators $\{P_{j}\}$, then
$\Theta(s,\sigma)$ is an orthogonal measurement on $l_{2}(Q)$
specified by projection operators $\{P_{j}^{'}\}\cup \{I_{1}\}$,
and $\delta(s,\sigma)=(s_{j},d_{j})$ if
$\delta_{2}(s,\sigma)=(s_{j},d_{j})$, where $P_{j}^{'}$ extend
$P_{j}$ to $l_{2}(Q)$ by defining $P_{j}^{'}|q_{1,i}\rangle=0$ for
$0\leq i\leq n_{1}$.
\end{enumerate}
\end{enumerate}
In terms of the 2qcfa $M$ constructed above, for any
$x\in\Sigma^{*}$, we have:
\begin{itemize}
\item If $x\in L_{1}\cap L_{2}$, then $M$ accepts $x$ with
probability at least

$(1-\epsilon_{1})(1-\epsilon_{2})=1-(\epsilon_{1}+\epsilon_{2}-\epsilon_{1}\epsilon_{2})$.

\item If $x\not\in L_{1}$, then $M$ rejects $x$ with probability
at least $1-\epsilon_{1}$.

\item If $x\in L_{1}$ but $x\not\in L_{2}$, then $M$ rejects $x$
with probability at least $(1-\epsilon_{1})(1-\epsilon_{2})$.
\end{itemize}
\end{proof}

By means of the proof of Theorem 1, we have the following
corollaries 1 and 2.

{\bf Corollary 1.} If languages $L_{1}$ and $L_{2}$ are recognized
by 2qcfa's $M_{1}$ and $M_{2}$ with one-sided error probabilities
$\epsilon_{1},\epsilon_{2}\in [0,\frac{1}{2})$ in polynomial time,
respectively, then $L_{1}\cap L_{2}$ is recognized by some 2qcfa
$M$ with one-sided error probability
$\epsilon=\max\{\epsilon_{1},\epsilon_{2}\}$ in polynomial time,
that is, for any input string $x$,
\begin{itemize}
\item if $x\in L_{1}\cap L_{2}$, then $M$ accepts $x$ with
certainty;

\item if $x\not\in L_{1}$, then $M$ rejects $x$ with probability
at least $1-\epsilon_{1}$;

\item if $x\in L_{1}$ but $x\not\in L_{2}$, then $M$ rejects $x$
with probability at least $1-\epsilon_{2}$.
\end{itemize}

{\bf Example 1.} We recall that non-regular language $L_{=}=\{x\in
\{a,b\}^{*}|\#_{x}(a)=\#_{x}(b)\}$. For non-regular language
$L_{=}(pal)=\{y=xx^{R}|x\in L_{=}\}$, we can clearly check that
$L_{=}(pal)=L_{=}\cap L_{pal}$. Therefore, by applying Corollary
1, we obtain that $L_{=}(pal)$ is recognized by some 2qcfa with
one-sided error probability $\epsilon$, since both $L_{=}$ and
$L_{pal}$ are recognized by 2qcfa's with one-sided error
probability $\epsilon$ [6], where $\epsilon$ can be given
arbitrarily small.

{\bf Corollary 2.} If $L_{1}\in 2QCFA(poly-time)$, $L_{2}\in
2QCFA(poly-time)$, then
\begin{enumerate}
\item $QS_{L_{1}\cap L_{2}}\leq QS_{L_{1}}+QS_{L_{2}}$;

\item $CS_{L_{1}\cap L_{2}}\leq CS_{L_{1}}+CS_{L_{2}}+QS_{L_{1}}$.
\end{enumerate}

Similar to Theorem 1, we can obtain the union operation of
2qcfa's.

{\bf Theorem 2.} If $L_{1}\in 2QCFA_{\epsilon_{1}}(poly-time)$ and
$L_{2}\in 2QCFA_{\epsilon_{2}}(poly-time)$ for
$\epsilon_{1},\epsilon_{2}\geq 0$, then $L_{1}\cup L_{2}\in
2QCFA_{\epsilon}(poly-time)$ with
$\epsilon=\epsilon_{1}+\epsilon_{2}-\epsilon_{1}\epsilon_{2}$.

\begin{proof} The idea is similar to the proof of Theorem 1. Let
$L_{i}$ be accepted by 2qcfa's $M_{i}$ with error probabilities
$\epsilon_{i}$ $(i=1,2)$. Then we construct a 2qcfa $M$ as the way
in Theorem 1, that is to say, we use $M$ firstly to simulate
$M_{1}$. If $M_{1}$ accepts, then $M$ also accepts; otherwise, $M$
continues to simulate $M_{2}$, and the accepting or rejecting of
$M$ depends on $M_{2}$. The process is more clearly described as
follows.

{\it For input string $x$, $M_{1}$ and $M_{2}$ with initial
quantum state $|q_{1,0}\rangle$ and  $|q_{2,0}\rangle$ as well as
classical state $s_{1,0}$ and $s_{2,0}$, respectively; also, $M$
has initial quantum state $|q_{1,0}\rangle$ and classical state
$s_{1,0}$. $M$ firstly simulate $M_{1}$. If $M_{1}$ accepts, then
$M$ accepts; if $M_{1}$ rejects, then $M$ becomes quantum state
$|q_{2,0}\rangle$ and classical state $s_{2,0}$, and continues to
simulate $M_{2}$. If $M_{2}$ rejects, then also $M$ rejects;
otherwise $M$ accepts as $M_{2}$ does.}

Similarly to Theorem 1, for any $x\in\Sigma^{*}$, we have:
\begin{itemize}
\item If $x\in L_{1}$, then $M$ accepts $x$ with probability at
least $1-\epsilon_{1}$.

\item If $x\not\in L_{1}$, but $x\in L_{2}$, then $M$ accepts $x$
with probability at least $(1-\epsilon_{1})(1-\epsilon_{2})$.

\item If $x\not\in L_{1}$ and $x\not\in L_{2}$, then $M$ rejects
$x$ with probability at least $(1-\epsilon_{1})(1-\epsilon_{2})$.
\end{itemize}
Since the specific process is analogous to Theorem 1, we leave the
details out here.

\end{proof}

Due to the proof of Theorem 2, we also have the following
corollary.

{\bf Corollary 3.} If languages $L_{1}$ and $L_{2}$ are recognized
by 2qcfa's $M_{1}$ and $M_{2}$ with one-sided error probabilities
$\epsilon_{1},\epsilon_{2}\in [0,\frac{1}{2})$ in polynomial time,
respectively, then there exists 2qcfa $M$ such that $L_{1}\cup
L_{2}$ is recognized by 2qcfa $M$ with error probability at most
$\epsilon_{1}+\epsilon_{2}-\epsilon_{1}\epsilon_{2}$ in polynomial
time, that is, for any input string $x$,
\begin{itemize}
\item if $x\in L_{1}$, then $M$ accepts $x$ with certainty;

\item if $x\not\in L_{1}$, but $x\in L_{2}$, then $M$ accepts $x$
with probability at least $1-\epsilon_{1}$;

\item if $x\not\in L_{1}$ and $x\not\in L_{2}$, then $M$ rejects
$x$ with probability at least $(1-\epsilon_{1})(1-\epsilon_{2})$.
\end{itemize}

Similar to Corollary 2, we have:

{\bf Corollary 4.} If $L_{1}\in 2QCFA(poly-time)$, $L_{2}\in
2QCFA(poly-time)$, then
\begin{itemize}
\item $QS_{L_{1}\cup L_{2}}\leq QS_{L_{1}}+QS_{L_{2}}$;

\item $CS_{L_{1}\cup L_{2}}\leq CS_{L_{1}}+CS_{L_{2}}+QS_{L_{1}}$.
\end{itemize}

{\bf Example 2.} As indicated in Remark 2,
$L_{eq}(k,a)=\{a^{kn}b^{n}|n\in {\bf N}\}$ and
$L_{eq}(k,b)=\{b^{n}a^{n}|n\in {\bf N}\}$ are recognized by
2qcfa's with one-sided error probabilities (as demonstrated by
Ambainis and Watrous [6], these error probabilities can be given
arbitrarily small) in polynomial time. Therefore, by using
Corollary 3, we have that for any $m\in {\bf N}$,
$\cup_{k=1}^{m}L_{eq}(k,a)$ and $\cup_{k=1}^{m}L_{eq}(k,b)$ are
recognized by 2qcfa's with error probabilities in
$[0,\frac{1}{2})$ in polynomial time.

For language $L$ over alphabet $\Sigma$, the complement of $L$ is
$L^{c}=\Sigma^{*}\backslash L$.  For the class of languages
recognized by 2qcfa's with bounded error probabilities, the unary
complement operation is also closed.

{\bf Theorem 3.} If $L\in 2QCFA_{\epsilon}(poly-time)$ for error
probability $\epsilon$, then $L^{c}\in
2QCFA_{\epsilon}(poly-time)$.

\begin{proof} Let 2qcfa
$M=(Q,S,\Sigma,\Theta,\delta,q_{0},s_{0},S_{acc},S_{rej})$ accept
$L$ with error probability $\epsilon\in [0,\frac{1}{2})$. Then we
can construct 2qcfa $M^{c}$ only by exchanging the classical
accepting and rejecting states in $M$, that is,
$M^{c}=(Q,S,\Sigma,\Theta,\delta,q_{0},s_{0},S_{acc}^{c},S_{rej}^{c})$
where $Q,S,\Sigma,\Theta,\delta,q_{0},s_{0}$ are the same as those
in $M$, and, $S_{acc}^{c}=S_{rej}$, $S_{rej}^{c}=S_{acc}$.
Clearly, $L^{c}$ is accepted by $M^{c}$ with error probability
$\epsilon$.

\end{proof}

From the proof of Theorem 3 it follows Corollary 5.

{\bf Corollary 5.} If $L\in 2QCFA(poly-time)$, then
\begin{itemize}
\item $QS_{L^{c}}= QS_{L}$; \item  $CS_{L^{c}}= CS_{L}$.
\end{itemize}

{\bf Example 3.} For non-regular language $L_{=}$, its complement
$L_{=}^{c}=\{x\in \{a,b\}^{*}|\#_{x}(a)\not=\#_{x}(b)\}$ is
recognized by 2qcfa with bounded error probability in polynomial
expected time, by virtue of Remark 3 and Theorem 3.

For language $L$ over alphabet $\Sigma$, the reversal of $L$ is
$L^{R}=\{x^{R}| x\in L\}$ where $x^{R}$ is the reversal of $x$,
i.e., if $x=\sigma_{1}\sigma_{2}\ldots\sigma_{n}$ then
$x^{R}=\sigma_{n}\sigma_{n-1}\ldots\sigma_{1}$. For
$2QCFA_{\epsilon}(poly-time)$ with $\epsilon\in [0,1/2)$, the
reversal operation is closed.

{\bf Theorem 4.} If $L\in  2QCFA_{\epsilon}(poly-time)$, then
$L^{R}\in  2QCFA_{\epsilon}(poly-time)$.

\begin{proof} Let $L$ be recognized by a 2qcfa $M$ with error
probability $\epsilon\in [0,\frac{1}{2})$. Then we can construct a
2qcfa $M^{c}$ simulate $M$ from the converse direction of the tape
head moving. More specifically, suppose
$M=(Q,S,\Sigma,\Theta,\delta,q_{0},s_{0},S_{acc},S_{rej})$. Then,
we construct
$M^{R}=(Q^{R},S^{R},\Sigma,\Theta^{R},\delta^{R},q_{0}^{R},s_{0}^{R},S_{acc}^{R},S_{rej}^{R})$
where $Q^{R}=Q\cup\{q_{0}^{R}\}$, $S^{R}=S\cup \{s_{0}^{R}\}$ with
$q_{0}^{R}\not\in Q$, $s_{0}^{R}\not\in S$, $S_{acc}^{R}=S_{acc}$,
$S_{rej}^{R}=S_{rej}$,
 $\Theta^{R}$ and
$\delta^{R}$ are defined as follows.
\begin{enumerate}
\item For $\sigma\in\Sigma\cup\{\cent\}$,
$\Theta^{R}(s_{0}^{R},\sigma)=I$, where $I$ is identity operator
on $l_{2}(Q^{R})$, $\delta^{R}(s_{0}^{R},\sigma)=(s_{0}^{R},1)$;
and $\Theta^{R}(s_{0}^{R},\$)=U(q_{0}^{R},q_{0})$, where
$U(q_{0}^{R},q_{0})$ is a unitary operator on $l_{2}(Q^{R})$
satisfying $U(q_{0}^{R},q_{0})|q_{0}^{R}\rangle=|q_{0}\rangle$,
and $\delta^{R}(s_{0}^{R},\$)=(s_{0},0)$.

\item For $s\in S$, $\sigma\in \Sigma\cup \{\cent, \$\}$, if
$\Theta(s,\sigma)$ is a unitary operator on $l_{2}(Q)$, then
$\Theta^{R}(s,\sigma)$ is also unitary operator on $l_{2}(Q^{R})$
by extending $\Theta(s,\sigma)$ with
$\Theta^{R}(s,\sigma)|q_{0}^{R}\rangle=|q_{0}^{R}\rangle$ and
$\Theta^{R}(s,\sigma)|\phi\rangle=\Theta(s,\sigma)|\phi\rangle$
for $|\phi\rangle\in l_{2}(Q)$, and
$\delta^{R}(s,\sigma)=(s^{'},-d)$ if $\delta(s,\sigma)=(s^{'},d)$.

\item For $s\in S$, $\sigma\in \Sigma\cup \{\cent, \$\}$, if
$\Theta(s,\sigma)$ is an orthogonal measurement on $l_{2}(Q)$
described by projectors $\{P_{j}\}$, then $\Theta^{R}(s,\sigma)$
is also an orthogonal measurement on $l_{2}(Q^{R})$ described by
projectors $\{P_{j}^{'}\}\cup\{I_{2}\}$, where $P_{j}^{'}$ extend
$P_{j}$ to $l_{2}(Q^{R})$ by defining
$P_{j}^{'}|q_{0}^{R}\rangle=0$, and projection operator $I_{2}$
mapping to $l_{2}(\{q_{0}^{R}\})$.
\end{enumerate}
Then, in terms of the 2qcfa $M^{R}$ constructed above, $M^{R}$
accepts $L^{R}$ with bounded error probability $\epsilon$.

\end{proof}

By means of the proof of Theorem 4 we clearly obtain the following
corollary.

{\bf Corollary 6.} If $L\in 2QCFA(poly-time)$, then
\begin{itemize}
\item $QS_{L}-1\leq QS_{L^{R}}\leq QS_{L}+1$; \item $CS_{L}-1\leq
CS_{L^{R}}\leq CS_{L}+1$.
\end{itemize}

For languages $L_{1}$ and $L_{2}$ over alphabets $\Sigma_{1}$ and
$\Sigma_{2}$, respectively, the catenation of $L_{1}$ and $L_{2}$
is
$L_{1}L_{2}=\{x_{1}x_{2}|x_{1}\in\Sigma_{1},x_{2}\in\Sigma_{2}\}$.
We do not know whether or not the catenation operation in
$2QCFA_{\epsilon}$ is closed, but under certain condition we can
prove that the catenation of two languages in $2QCFA_{\epsilon}$
is closed.

{\bf Theorem 5.} Let $L_{i}\in 2QCFA_{\epsilon}$, and
$\Sigma_{1}\cap\Sigma_{2}=\emptyset$ where $\Sigma_{i}$ are
alphabets of $L_{i}$ $(i=1,2)$. Then the catenation $L_{1}L_{2}$
of $L_{1}$ and $L_{2}$ is also recognized by a 2qcfa with error
probability at most
$\epsilon=\epsilon_{1}+\epsilon_{2}-\epsilon_{1}\epsilon_{2}$.

\begin{proof} Let $\varepsilon$ denote empty string. We may consider four cases:
\begin{enumerate}
\item $\varepsilon\not\in L_{1}\cup L_{2}$;
\item
$\varepsilon\not\in L_{2}$ but $\varepsilon\in L_{1}$;
 \item
$\varepsilon\not\in L_{1}$ but $\varepsilon\in L_{2}$;
 \item
$\varepsilon\in L_{1}\cap L_{2}$.
\end{enumerate}
Here we only prove case 1, since the other cases are similar.

Suppose that $L_{i}$ are recognized by 2qcfa's
$M_{i}=(Q_{i},S_{i},\Sigma_{i},\Theta_{i},\delta_{i},q_{i,0},s_{i,0},S_{i,acc},S_{i,rej})$,
with error probabilities $\epsilon_{i}$ $(i=1,2)$. Then we
construct 2qcfa $M$ accepting $L_{1}L_{2}$ with error probability
$\epsilon=\epsilon_{1}+\epsilon_{2}-\epsilon_{1}\epsilon_{2}$.
Firstly we let $M$ check whether or not the input is the form of
$xy\in \Sigma_{1}^{+}\Sigma_{2}^{+}$ where $\Sigma_{i}^{+}$ denote
the set of all non-empty strings over $\Sigma_{i}$; otherwise $M$
rejects the input immediately. Then let $M$ simulate $M_{1}$, and,
as soon as $M_{1}$ meets an input symbol not in $\Sigma_{1}$,
$M_{1}$ views this input symbol as $\$$. Therefore, if $M_{1}$
rejects the first part of input string, then $M$ rejects the input
string; otherwise, $M$ continues to compute the second part of the
input string by simulating $M_{2}$, and, therefore, the results of
rejecting and accepting of $M$ further depend on $M_{2}$. Hence,
the computing process of $M$ is roughly as follows.

{\it For input string $x$, $M$ checks whether $x$ is the form in
$\Sigma_{1}^{+}\Sigma_{2}^{+}$. If it is not such a form, then $M$
rejects it; otherwise $M$ continues to simulate $M_{2}$, for
resulting in the accepting and rejecting probabilities. }

More formally, let
$M_{i}=(Q_{i},S_{i},\Sigma_{i},\Theta_{i},\delta_{i},q_{i,0},s_{i,0},S_{i,acc},S_{i,rej})$
$(i=1,2)$. Then
$M=(Q,S,\Sigma,\Theta,\delta,q_{0},s_{0},S_{acc},S_{rej})$, where:
\begin{itemize}
\item $Q=Q_{1}\cup Q_{2}$, \item $S=S_{1}\cup
S_{2}\cup\{s_{0},s_{1},s_{2},s_{3}\}$ with
$\{s_{0},s_{1},s_{2},s_{3}\}\cap (S_{1}\cup S_{2})=\emptyset$,
\item $q_{0}=q_{1,0}$, \item $S_{acc}=S_{acc,2}$, \item
$S_{rej}=S_{1,rej}\cup S_{2,rej}\cup \{s_{2}\}$,
\end{itemize}
and $\Theta$ and $\delta$ are defined as follows.
\begin{enumerate}
\item Firstly, let $M$ check the form of the input string,
\begin{enumerate}
\item for any $\sigma\in \Sigma_{1}\cup\{\cent\}$,
$\Theta(s_{0},\sigma)=I$, where $I$ is identity operator on
$l_{2}(Q)$, $\delta(s_{0},\sigma)=(s_{0},1)$;

\item for any $\sigma\in \Sigma_{1}$, $\Theta(s_{1},\sigma)=I$,
$\delta(s_{1},\sigma)=(s_{2},0)$, where $s_{2}\in S_{rej}$;

\item for  any $\sigma\in \Sigma_{2}$, $\Theta(s_{0},\sigma)=I$,
$\delta(s_{0},\sigma)=(s_{1},1)$;

\item for $\sigma=\$$, $\Theta(s_{1},\$)=I$,
$\delta(s_{1},\$)=(s_{3},-1)$;

\item for $\sigma\in \Sigma_{1}\cup\Sigma_{2}$,
$\Theta(s_{3},\sigma)=I$, $\delta(s_{3},\sigma)=(s_{3},-1)$; \item
for $\sigma=\cent$, $\Theta(s_{3},\sigma)=I$,
$\delta(s_{3},\sigma)=(s_{1,0},0)$.
\end{enumerate}

\item Secondly, let $M$ simulate $M_{1}$. For $\sigma\in
\Sigma_{1}\cup\{\cent\}$, $s\in S_{1}$,
\begin{enumerate}
\item if $\Theta_{1}(s,\sigma)$ is a unitary operator on
$l_{2}(Q_{1})$, then $\Theta(s,\sigma)$ is also a unitary operator
on $l_{2}(Q)$ by extending $\Theta_{1}(s,\sigma)$ with
$\Theta(s,\sigma)|q\rangle=|q\rangle$ for $q\in Q_{2}$ and
$\Theta(s,\sigma)|\phi\rangle=\Theta_{1}(s,\sigma)|\phi\rangle$
for $|\phi\rangle\in l_{2}(Q_{1})$, and
$\delta(s,\sigma)=\delta_{1}(s,\sigma)$;

\item if $\Theta_{1}(s,\sigma)$ is an orthogonal measurement
described by projectors $\{P_{j}\}$, then $\Theta(s,\sigma)$ is
also an orthogonal measurement specified by projection operators
$\{P_{j}^{'}\}\cup \{I_{2}\}$, where $P_{j}^{'}$ are the
extensions of $P_{j}$ to $l_{2}(Q)$ by defining
$P_{j}^{'}|q\rangle=0$ for $q\in Q_{2}$, and $I_{2}$ is identity
operator on $l_{2}(Q_{2})$, and $I_{2}|q\rangle=0$ for $q\in
Q_{1}$; on the other hand, the definition of $\delta(s,\sigma)$ is
in terms of $\delta_{1}(s,\sigma)$, i.e., $\delta(s,\sigma)$ maps
the measuring result of $P_{j}^{'}$ to the same element as
$\delta_{1}(s,\sigma)(j)$, and $\delta(s,\sigma)$ maps the
measuring result of $I_{2}$ to any classical state and direction
(indeed, before measuring, the quantum superposition state does
not include $Q_{2}$, and, therefore, the probability of obtaining
measuring result by performing operator $I_{2}$ is zero).
\end{enumerate}
\item For $\sigma\in \Sigma_{2}$, $s\in S_{1}$,
$\Theta(s,\sigma)=\Theta_{1}(s,\$)$ and
$\delta(s,\sigma)=\delta_{1}(s,\$)$.

\item For $s\in S_{1,acc}$,
\begin{enumerate}
\item if $\sigma\in\Sigma_{1}\cup\{\cent\}$, then
$\Theta(s,\sigma)=I$, where $I$ is identity operator on
$l_{2}(Q)$, and $\delta(s,\sigma)=(s,1)$; \item if
$\sigma\in\Sigma_{2}$, then $\Theta(s,\sigma)=I$, where $I$ is
identity operator on $l_{2}(Q)$, and
$\delta(s,\sigma)=(s_{2,0},-1)$.
\end{enumerate}
\item For
$\sigma\in \Sigma_{1}$,
\begin{enumerate}
\item if $\Theta_{2}(s_{2,0},\cent)$ is a unitary operator on
$l_{2}(Q_{2})$, then $\Theta(s_{2,0},\sigma)$ is also a unitary
operator on $l_{2}(Q)$ by directly extending
$\Theta_{2}(s_{2,0},\cent)$, and
$\delta(s_{2,0},\sigma)=\delta_{2}(s_{2,0},\cent)$;

\item if $\Theta_{2}(s_{2,0},\cent)$ is an orthogonal measurement
on $l_{2}(Q_{2})$ described by projectors $\{P_{j}\}$, then
$\Theta(s_{2,0},\sigma)$ is also an orthogonal measurement on
$l_{2}(Q)$ specified by $\{P_{j}^{'}\}\cup \{I_{1}\}$ where
$P_{j}^{'}$ are the extensions of $P_{j}$ to $l_{2}(Q)$ by
defining $P_{j}^{'}|q\rangle=0$ for $q\in Q_{1}$, and $I_{1}$ is
defined as $I_{1}|q\rangle=\left\{\begin{array}{ll}|q\rangle,&
q\in
Q_{1},\\
0, & q\in Q_{2},
\end{array}\right.$
and, as above,
$\delta(s_{2,0},\sigma)(j)=\delta(s_{2,0},\cent)$$(j)$.
\end{enumerate}
\item For $\sigma\in \Sigma_{2}\cup \{\$\}$ and $s\in S_{2}$,
$\Theta(s,\sigma)$ and $\delta(s,\sigma)$ are defined by means of
$\Theta_{2}(s,\sigma)$ and $\delta_{2}(s,\sigma)$ in the light of
Case 5 above.
\end{enumerate}

According to the 2qcfa $M$ specified above, for any $x\in
(\Sigma_{1}\cup\Sigma_{2})^{*}$, we have:
\begin{itemize}
\item If $x$ is not in $\Sigma_{1}^{+}\Sigma_{2}^{+}$, then $x$ is
rejected with certainty.

\item If $x$ is in $\Sigma_{1}^{+}\Sigma_{2}^{+}$, say
$x=x_{1}x_{2}$ for $x_{i}\in \Sigma_{i}^{+}$ $(i=1,2)$, then a) if
$x_{1}\not\in L_{1}$, then $x$ is rejected with probability at
least $1-\epsilon_{1}$, b) and if $x_{1}\in L_{1}$ and
$x_{2}\not\in L_{2}$, then $x$ is rejected with probability at
least $(1-\epsilon_{1})(1-\epsilon_{2})$.

\item If $x\in L_{1}L_{2}$, then $x$ is accepted by $M$ with
probability at least $(1-\epsilon_{1})(1-\epsilon_{2})$.
\end{itemize}

\end{proof}

From Theorem 5 it follows the following corollary.

{\bf Corollary 7.} Let languages $L_{i}$ over alphabets
$\Sigma_{i}$ be recognized by 2qcfa's with one-sided error
probabilities $\epsilon_{i}$ $(i=1,2)$ in polynomial time. If
$\Sigma_{1}\cap \Sigma_{2}=\emptyset$, then the catenation
$L_{1}L_{2}$ is recognized by some 2qcfa with one-sided error
probability $\max\{\epsilon_{1},\epsilon_{2}\}$, in polynomial
time.

{\bf Remark 4.} As indicated in Remark 1, the catenation,
$\{a^{n}b_{1}^{n}a^{m}b_{2}^{m}|n,m\in {\bf N}\}$, of
$L_{eq}^{(1)}=\{a^{n}b_{1}^{n}|n\in {\bf N}\}$ and
$L_{eq}^{(2)}=\{a^{n}b_{2}^{n}|n\in {\bf N}\}$, can also be
recognized by some 2qcfa with one-sided error probability
$\epsilon$ in polynomial time, where $\epsilon$ can be arbitrarily
small. Therefore, in Theorem 5, the condition of
$\Sigma_{1}\cap\Sigma_{2}=\emptyset$ is not necessary.

\section*{4. Concluding remarks}

2qcfa's were introduced by Ambainis and Watrous [6], and this kind
of computing models with classical tape heads is more restricted
than the usual 2qfa's [23], but it is still more powerful than
2pfa's. As a continuation of [6], in this note, we have dealt with
a number of operation properties of 2qcfa's. We proved that the
Boolean operations (intersection, union, and complement) and the
reversal operation of the class of languages recognized by 2qcfa's
with error probabilities are closed; as corollaries, we showed
that the intersection, complement, and reversal operations in the
class of languages recognized by 2qcfa's with one-sided error
probabilities (in $[0,\frac{1}{2})$) are closed. Furthermore, we
verified that the catenation operation in the class of languages
recognized by 2qcfa's with error probabilities is closed under
certain restricted condition (this result also holds for the case
of one-sided error probabilities belonging to $[0,\frac{1}{2})$).
As well, the numbers of states of these 2qcfa's for the above
operations were presented, and some examples were included for an
application of the derived results. For instance, $\{xx^{R}|x\in
\{a,b\}^{*},\#_{x}(a)=\#_{x}(b)\}$ was shown to be recognized by
2qcfa with one-sided error probability $0\leq \epsilon<
\frac{1}{2}$ in polynomial time.

These operation properties presented may apply to 2qfa's [23], but
the unitarity should be satisfied in constructing 2qfa's, and,
therefore, more technical methods are likely needed or we have to
add some restricted conditions (for example, we may restrict the
initial state not to be entered again). On the other hand, in
Corollaries 2 and 4, the lower bounds need be further fixed. We
would like to further consider them in the future.

\section*{Acknowledgement}

I would like to thank Dr. Tomoyuki Yamakami for helpful discussion
regarding quantum automata.

\end{document}